\documentclass[fleqn,usenatbib]{mnras}

\usepackage[T1]{fontenc}
\usepackage{ae,aecompl}
\usepackage[export]{adjustbox}

\usepackage{graphicx}	
\usepackage{amsmath}	
\usepackage{amssymb}	

\author[P.~Frick et al.]
       {P.~Frick$^1$,  D.~Sokoloff$\,^{1,3,4}$,  R.~Stepanov$^{1,6}$, V.~Pipin$^2$, and I.~Usoskin$^5$\\
$^1$ Institute of Continuous Media Mechanics,
Korolyov str.~1, 614013 Perm, Russia \\
$^2$ Institute of solar-terrestrial physics, Irkutsk, Russia \\
$^3$ Department of Physics, Moscow University, 119899, Moscow, Russia \\
$^4$ IZMIRAN, Kaluzhskoe shasse, 4, Troitsl,  Moscow, 108840,  Russia \\
$^5$ University of Oulu, 90014 Oulu, Finland \\
$^6$ Perm National Research Polytechnic University, 614990 Perm, Russia}

\title{Spectral characteristic of mid-term quasi-periodicities in sunspots data
}

\date{Accepted XXX. Received 2019; in original form ZZZ}

\pubyear{2019}

\begin{document}
\label{firstpage}
\pagerange{\pageref{firstpage}--\pageref{lastpage}}
\maketitle

\begin{abstract}Numerous analyses suggest the existence of various quasi-periodicities in solar activity. 
The power spectrum of solar activity recorded in sunspot data is dominated by the $\sim$11-year quasi-periodicity, known as the Schwabe cycle. 
In the mid-term range (1 month\,--\,11 years) a pronounced variability known as a quasi-biennial oscillation (QBO) is widely discussed. 
In the shorter time scale a pronounced peak, corresponding to the synodic solar rotation period ($\sim$ 27 days) is observed.  
Here we revisited the mid-term solar variability in terms of statistical dynamic of fully turbulent systems, where solid arguments are required to accept an isolated dominant frequency in a continuous (smooth) spectrum. 
For that, we first undertook an unbiased analysis of the standard solar data, sunspot numbers { and the F10.7 solar radioflux index}, by applying a wavelet tool, which allows one to perform a frequency-time analysis of the signal.
Considering the spectral dynamics of solar activity cycle by cycle, we showed that no single periodicity can be separated, in a statistically significant manner, in the specified range of periods. We examine whether a model of solar dynamo can reproduce the mid-term oscillation pattern observed in solar data.  
We found that a realistically observed spectrum can be explained if small spatial (but not temporal) scales are effectively smoothed. 
This result is important because solar activity is a {\it global} feature,  although monitored via small-scale tracers like sunspots.
\end{abstract}


\section{Introduction}

Cyclic solar activity is mostly presented by the dominant famous $\sim$11-year cycle known as the Schwabe cycle. 
This cycle was discovered in the 19th century via a simple analysis of sunspot numbers, but it is apparent also in other tracers of solar activity. 
The origin of the cycle is believed to be associated with the solar dynamo action. 
More specifically, the Schwabe cycle is understood as a leading eigensolution of solar dynamo  equations. 
Using different magnetic tracers, solar physicists found that the cycle is formed by propagation of a wave of quasi stationary magnetic field and that it is in fact a 22-year magnetic cycle, which is observed as an 11-year one because the sunspot number is insensitive to the sign of magnetic field. 

Variability of solar activity is however much more complex than just the Schwabe cycle. 
Analyses of sunspot data undertaken by many researchers \citep[see for review, e.g.,][]{hathawayLR,usoskin_LR_17} suggested that various periods exist in solar variability. 
Some of these periods are longer than the Schwabe cycle (e.g., the Gleissberg cycle with a typical timescale of about 100 years) while others are shorter. 
Most discussed quasi-periodic cycles in the mid-term range are attributed to oscillations at about two-year timescale \citep[e.g.,][]{1995SoPh..161....1B},  called the quasi-biennial oscillation (QBO). 
It was shown later that QBO corresponds to a wide range of periods \citep[0.6\,--\,4 years according to][]{2014SSRv..186..359B} and behaves intermittently. 
QBO has also been referred to as intermediate- or mid-term quasi-periodicities, identified with particular spectral peaks.   
However, this definition obviously deserves further clarification and discussion. 
If accepted naively, it reads that the period of oscillation varies with a characteristic time, which is substantially larger than the nominal period. 
This is likely a case when the original definition is misleading, and new concepts are required. We note that this is not solely a problem of terminology, since sunspot data and their interpretation form a basis for solar dynamo studies, where periodicities are associated with eigen frequencies of solar dynamo.

Several physical mechanisms were proposed to explain the QBO \citep[e.g,][]{Fetal10, Zetal10, Setal13, Detal18}. 
\cite{Ietal19} demonstrated that, while some contemporary detailed dynamo models can yield oscillations similar to the QBO, others do not, which makes QBO a kind of test for different dynamo models. 
We suggest a revised physical explanation of the QBO phenomenon observed in the mid-term range of sunspots variations. 

From the point of view of large-scale dynamics, the Sun is a very complicated magneto-hydrodynamical system operating under large values of the control parameters (Reynolds, Grashof, Hartmann numbers), in which the largest (lowest in Fourier space) modes appear on the background of fully developed turbulent media. 
A similar problem of identification of quasi-stable oscillations in turbulent systems is typical for fluid dynamics. 

It is worth to note that the discovery of large-scale flow \citep[convective wind, see e.g.,][]{1981PNAS...78.1981K,1983PhyD....9..287B} in fully developed convective systems in a laboratory (notably at much more moderate scales and parameters than solar plasma) was recognized by the community with some skepticism. It took time to bring the study of the dynamics of large-scale circulation in a separate topic of Rayleigh-B\'enard convection  \citep{2009RvMP...81..503A}.
The first attempts to study the temporal dynamics of large-scale modes in a convective cell revealed a complex spectrum with a series of peaks \citep{1980JETPL..32..210B}. Similar features were detected in the space-time spectra (that is, the temporal spectra of the isolated spatial mode) of various turbulent hydrodynamic systems and suggested similarities with the data on solar-related periodicities \citep{Zimin1988b}. However, attempts to obtain reliable sequences in long-running (several weeks) laboratory experiments showed that as soon as statistics become reliable (the sample coverage greatly exceeds the characteristic times studied), the spectra become smooth and only one dominant frequency survives in some cases \citep{2001JFM...449..169N,2001PhRvL..87i4501Q,Vasiliev:IJHMT2016}.

In this paper we suggest, based on the background of turbulent and convection studies, that the mid-term (0.1\,--\,11 years) solar dynamics should be considered in term of statistical dynamics of fully turbulent systems. Otherwise convincing physical arguments or robust statistics are required to distinguish a dominant frequency in a continuous spectrum.

For that, we first undertake an unbiased analysis of the standard solar data, viz. sunspot number, considering the spectral dynamics of solar activity cycle by cycle. 
{ The results of the analysis were verified also with the F10.7 cm solar radioflux index.}
As a spectral tool we use a wavelet analysis, which occurs helpful in studies of solar variability \citep[e.g.,][]{1995ApJ...455..366L,Frick97}.  
We then apply a simplified non-axisymmetric model of solar dynamo to reproduce the observed spectral statistics in the mid-term range. 


\section{Data}

We based our analysis on sunspot activity as quantified in several indices. 
We analyzed the International total sunspot number \citep[ISN, v.2.0,][]{clette16}, which is based on the classical Wolf sunspot number series with correction of some apparent errors and re-calibration to Wolfer as the reference observer. Sunspot number represents a synthetic number being a combination of the weighted number of sunspot groups and the number of individual spots, corrected for the individual observer's quality factor.
Henceforth it is called the SN-series. The data is available from the SILSO database \url{http://www.sidc.be/silso/datafiles}.
 
We also considered the sunspot group number (notated  henceforth GN), which provides the number of sunspot groups visible on the solar disc at a given time. This series was obtained by \cite{usoskin16} using the active-day-fraction method from the  raw database of sunspot group observations \citep{vaquero16} for the period 1749\,--\,1995. 
This dataset is available at \url{ http://www.sidc.be/silso/DATA/GroupNumber/GNiu_d.txt}.

We also considered the sunspot area (notated  henceforth SA) as observed by the Royal Greenwich Observatory (RGO) for the period 1876\,--\,1976 and extended after 1976 by the USAF/NOAA data set scaled with the factor 1.4. Both hemispheric and global sunspot areas were used. The data set and its full description can be found at \url{https://solarscience.msfc.nasa.gov/greenwch.shtml}  \citep[see also][]{hathawayLR}.

{  In addition, we also considered the F10.7 solar index, which is solar radio flux in the wavelength 10.7 cm (2800 MHz) originated from the upper chromosphere / lower corona. It forms one of the longest directly measured (in contrast to synthetic sunspot data) solar indices and is available since the mid-20th century. 
The dataset was obtained from the OMNIWeb data service  (\url{https://omniweb.gsfc.nasa.gov}).
}

We used daily data, which is the best cadence for this type of data, but we appreciate that the sunspot's lifetime is longer (days\,--\,weeks for individual spots and months for an active region), so that the results of our analysis (especially at time scales comparable with the spot lifetime) can be determined by both the dynamo mechanism of cyclic solar magnetic activity and physics of sunspot/active-region evolution. 

\begin{figure}
\centering
\includegraphics[width=\columnwidth]{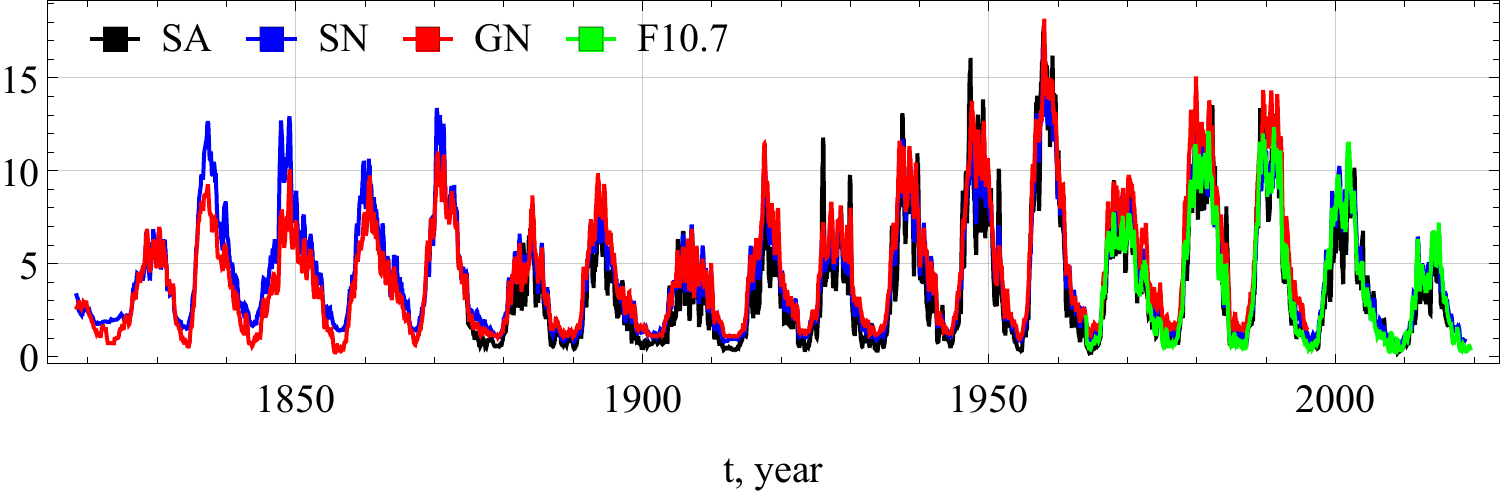}
\caption{Data sets used in this work: sunspot area (SA, black, normalized by a factor 226), sunspot number (SN, blue, normalized by 20.01), sunspot group number (GN, red, normalized by 20.01),  {the solar radio flux in the wavelength 10.7 cm (F10.7, green, normalized by 20.01).} }
\label{fig:data}
\end{figure}

The time series used here are shown in Fig.~\ref{fig:data}. We note that each data sets covers specific time intervals. As the data sets deals with slightly different kind of tracers, normalization is required to make the results of our analysis comparable. We use the following normalizations: SN, GN and F10.7 are normalized by a factor 20.01, and SA is normalized by 226.

\begin{figure}
\centering
\includegraphics[width=\columnwidth]{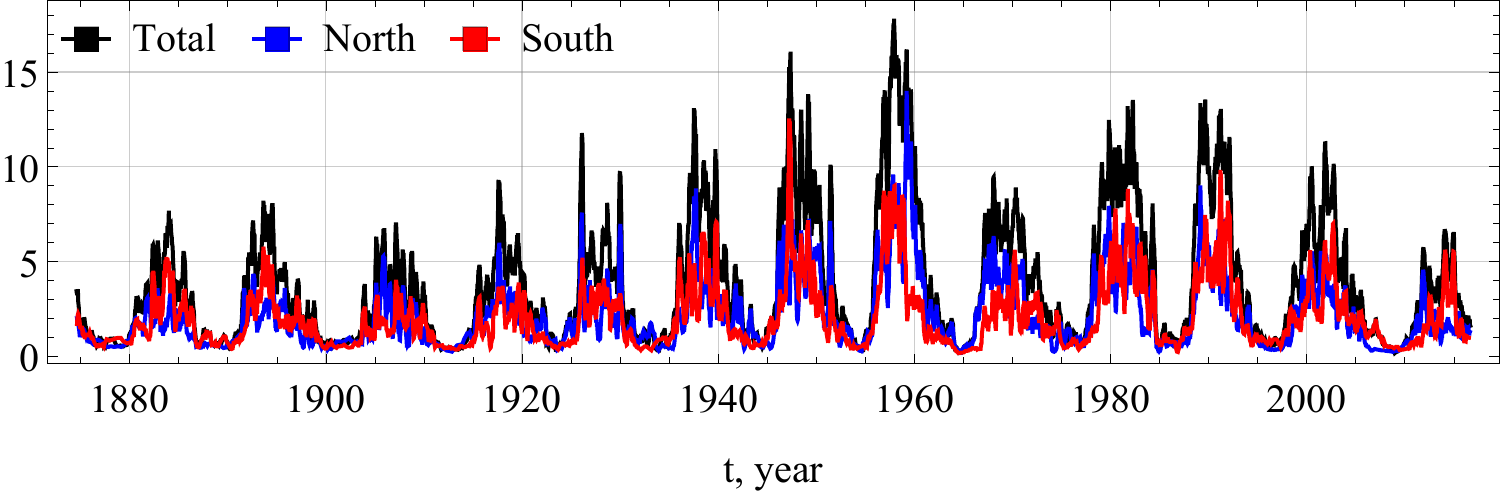}
\caption{Time series for sunspots area (SA), normalized by a factor 226: total (black), for the North hemisphere (blue), for the  South hemisphere (red).}
\label{fig:dataSN}
\end{figure}

Solar dynamo produces two 11-year periodic activity waves, one in each hemisphere. In principle, periodicity's under discussion might be associated with activity waves in particular hemisphere. To test this option we analyse additionally the data set for particular hemispheres separately. 
All used data sets provide indication of the hemisphere where a particular sunspot is located. As an example, we show in Fig.~\ref{fig:dataSN} time series of sunspot area separated by hemispheres in comparison with total data.

\section{Wavelets}

Our aim is to identify quasi-periodic components in a stochastic and noisy signal, considering that both the period and amplitude of oscillations may vary from cycle to cycle \citep{hathawayLR}. 
There is a suitable mathematical tool for such an analysis known as wavelet analysis  \citep[e.g.][]{Mallat2008}, which is a localized version of Fourier analysis: the analyzed signal is compared with a wave packet of various wavelengths centered at various time. Wavelets are successfully used to analyze sunspot data since the first study by \citet{1995ApJ...455..366L,1995CRASB.321..525N}. 

In general, various profiles of wave packets may be exploited, but here we use harmonic oscillations modulated by a Gaussian envelope (the so-called Morlet wavelet),
\begin{equation}\label{psi}
  \psi(t)=\sigma^{-1/2}e^{-(t/\sigma)^2} \left(e^{\imath t \omega }-e^{-\sigma ^2 \omega ^2 /4}\right),
\end{equation}
with $\omega = 2 \pi$ and $\sigma = 1$. This values provide appropriate  resolution in time and period estimates \citep{1999ApJ...510L.135S}. 

Then the wavelet transform is defined as
\begin{equation}\label{wc}
  W_\tau(t)=\tau^{-1} \int_{t_0}^{t_1} f(t')\, \psi^*\left(\frac{t-t'}{\tau}\right)  dt',
\end{equation}
and the global wavelet spectrum (corresponding generalization of the Fourier spectrum) is
\begin{equation}\label{Ea}
  E(\tau)=\tau \int_{t_0}^{t_1} |W_\tau(t)|^2 dt.
\end{equation}

Obtaining the global wavelet spectrum (\ref{Ea}) from 
the decomposition (\ref{wc}), we can also perform  integration over a specific time interval, say, 
a particular Schwabe cycle, to get the spectral 
characteristic of solar activity inside this 
cycle only.  
We stress here that this calculation requires the data being available beyond this particular cycle 
because (\ref{wc}) formally contains the 
integration over the whole database used. 
Having wavelet spectra for several Schwabe 
cycles we can estimate the standard deviation for the global wavelet spectrum as scattering of wavelet 
spectra of individual cycles. Below we separate 
one Schwabe cycle from the next one using the 
instants at which corresponding phase of $W_{11}$ (wavelet coefficient at $\tau=11$) crosses $\pi/2$.

The standard version of wavelet analysis faces 
problems  dealing with gaps and edges in data set. The 
problem can be partly resolved with 
so-called gaped wavelet algorithm suggested for stellar cyclic activity investigations  \citep{Frick1998}.

\section{Results of wavelets analysis}
\label{ResultsW}

\begin{figure}
\centering
\includegraphics[width=0.7\columnwidth]{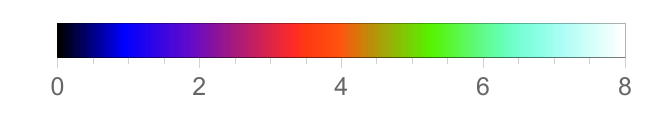}
\includegraphics[width=\columnwidth]{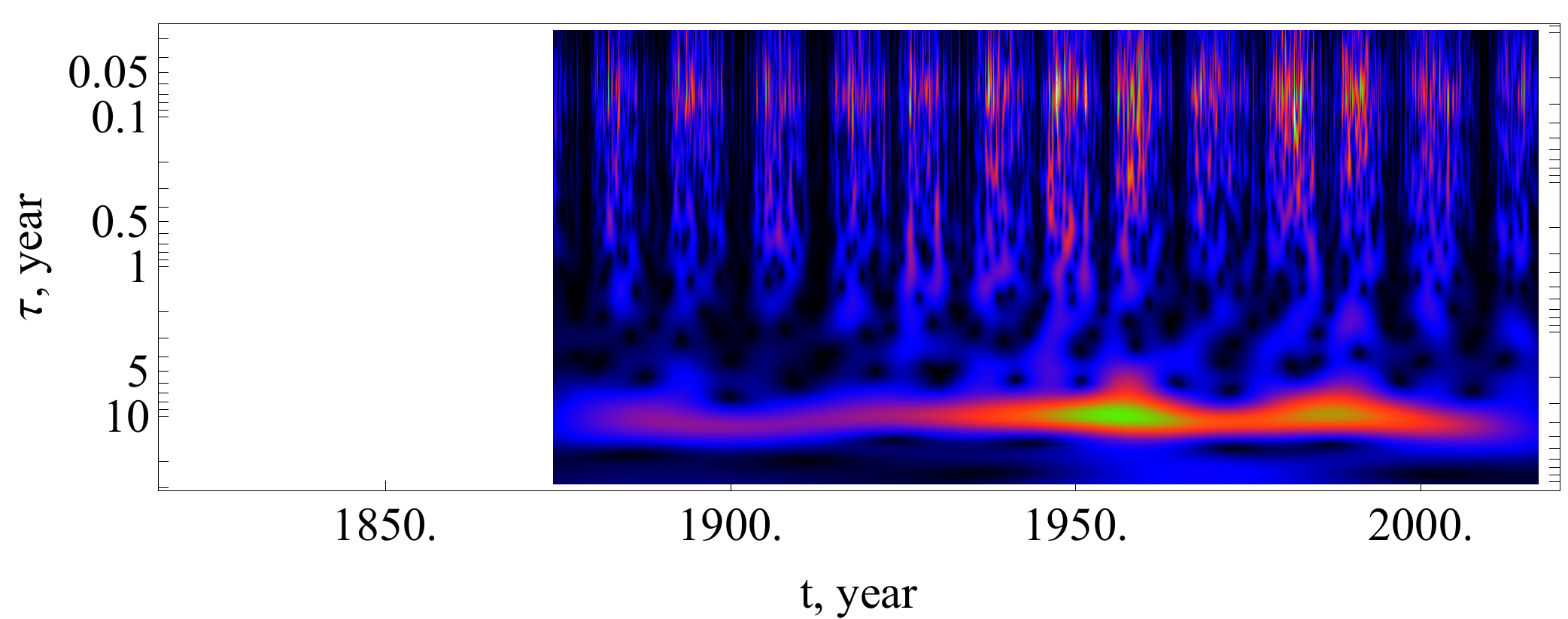}
\includegraphics[width=\columnwidth]{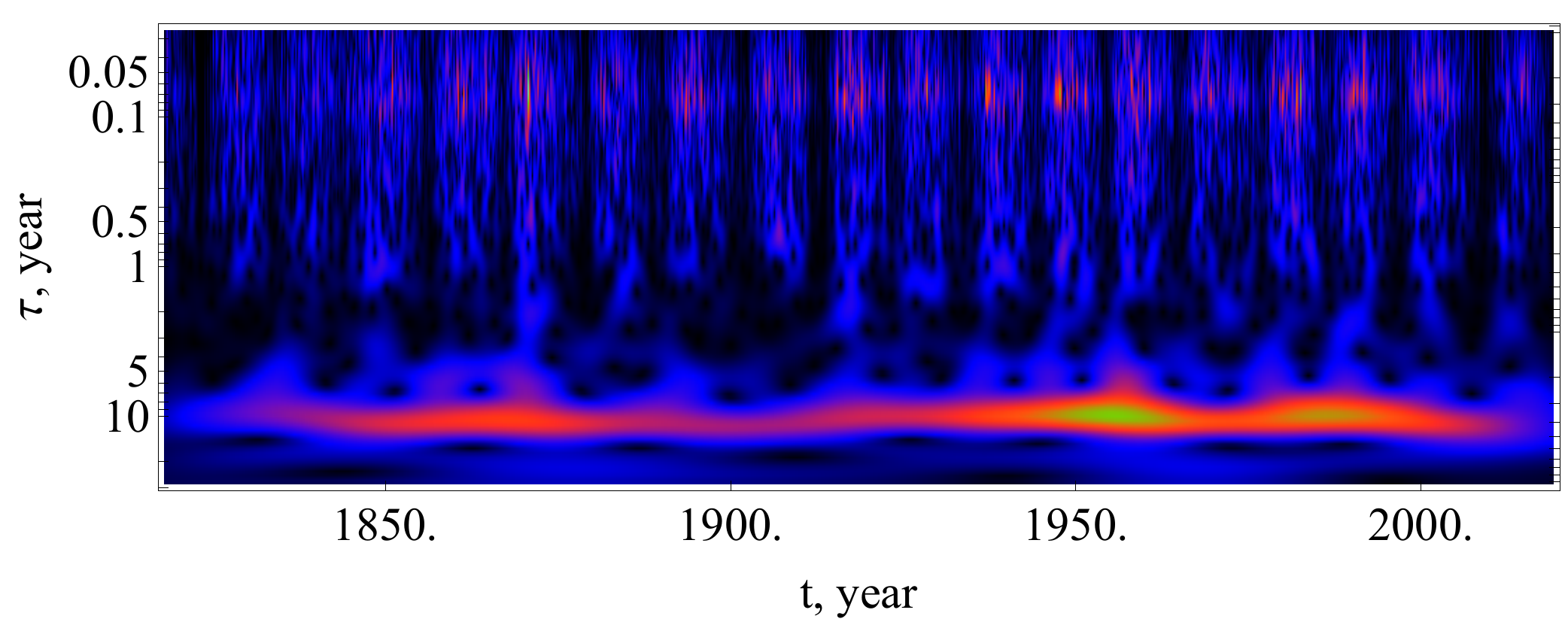}
\includegraphics[width=\columnwidth]{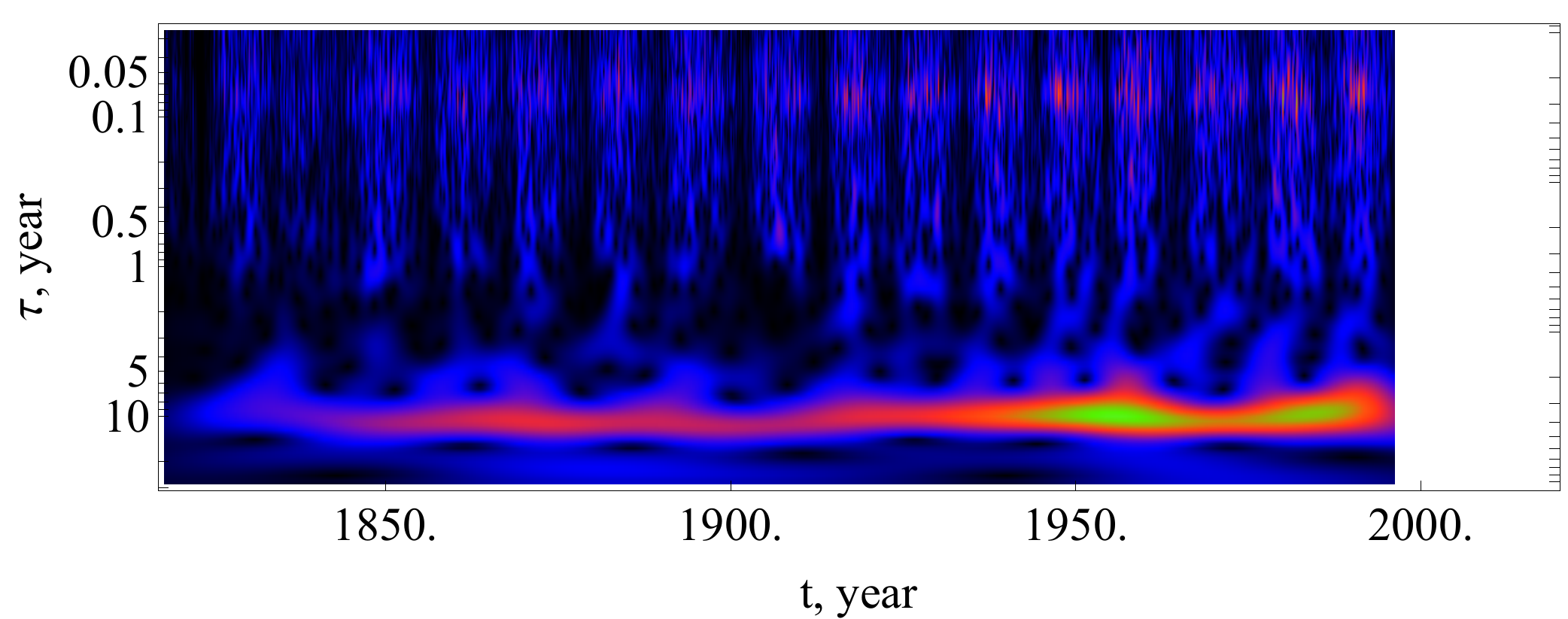}
\includegraphics[width=\columnwidth]{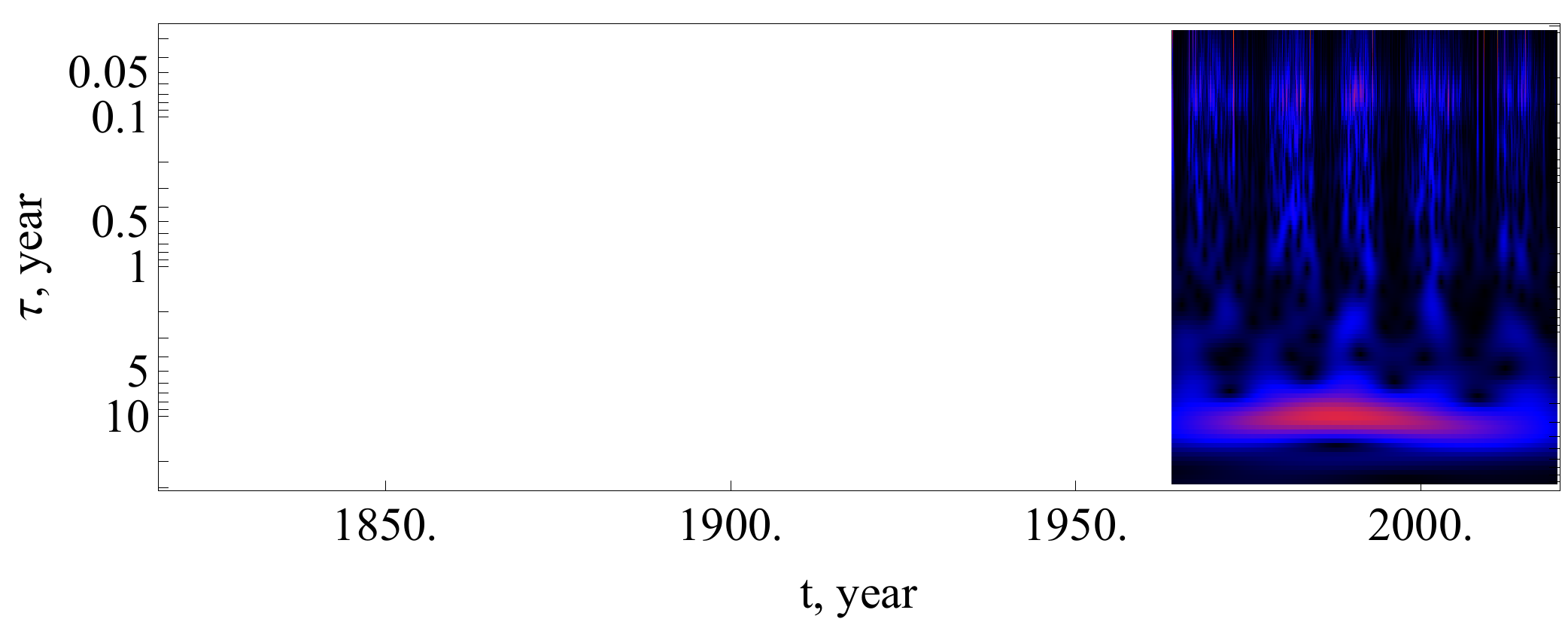}
\caption{Wavelet spectrograms for  SA, SN, GN and F10.7 (from top to bottom). Colours from blue to red correspond to the intensity of $|W_\tau (t)|$.}
\label{fig:maps}
\end{figure}

First we present wavelet spectrograms in the $\tau$-vs-$t$ coordinates, where colour corresponds to the modulus $|W_\tau (t)|$  (Fig.~\ref{fig:maps}). 
Two pronounced strips at $\tau_1 \approx 11$ years and $\tau_2 \approx 0.1$ year can be seen, corresponds to the nominal 11-year Schwabe cycle and the (synodic) solar rotation period, respectively.
Some isolated 'isles' of enhanced power can be also observed between $\tau_1$ and $\tau_2$, i.e. in the frequency/period domain under discussion. 

\begin{figure}
\centering
\includegraphics[width=\columnwidth]{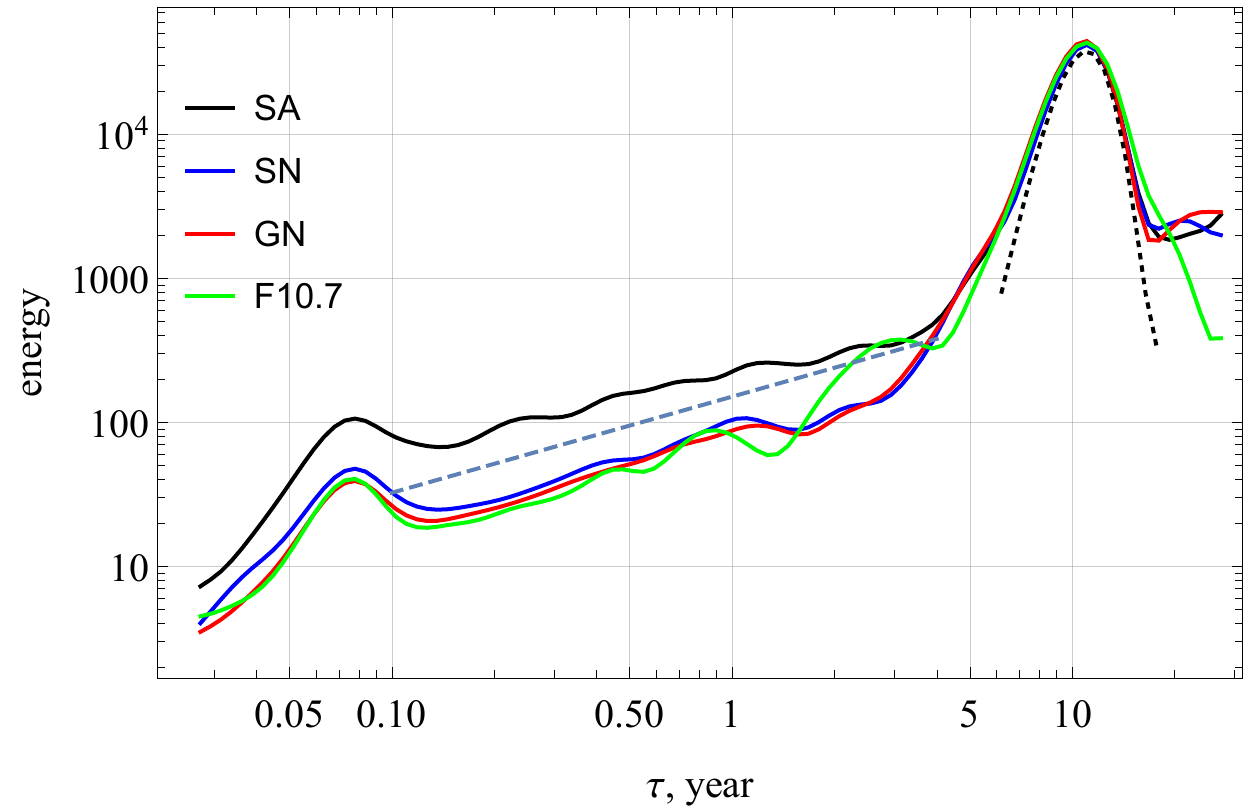}
\caption{Global wavelet spectra for different tracers: SA (black), SN (blue), GN (red) and  {F10.7 (green)}. 
The dashed line denotes the slope ${2/3}$. 
The dotted line denotes the response expected from a pure 11-year sinusoidal signal.}
\label{fig:total}
\end{figure}

Global wavelet spectra (meaning integral (\ref{Ea}) is taken over entire length of time series) are shown in Fig.~\ref{fig:total} and display two significant peaks at $\tau_1$ and $\tau_2$.
All {  solar activity indices} give very similar global spectra. 
We also calculated wavelet spectra for GN excluding small ($< 35\mu$sd) sunspot groups 
 to verify the robustness of the spectral shape.

The highest peak in the global spectrum is quite broad corresponds to the nominal 11-year Schwabe cycle. 
We note that, since wavelets have a finite spectral resolution, they typically yield a smooth peak even for a purely harmonic signal, 
 as illustrated by the dotted curve in Fig.~\ref{fig:total} which corresponds to a purely sinusoidal 11-year variation. 

The second peak is also broad and corresponds to the time scale of $\approx 27$ days, which is the solar synodic rotation period.
This peak is formed by long-living active regions, whose life time exceeds one solar rotation period, and thus they can recurrently
 contribute to solar activity as traced from Earth. 
Since new active regions are formed at different latitudes and random longitudes, the peak appears broad.
The ability to reproduce the rotational period solar-activity variations is tested below in the framework
of a solar dynamo model (Sect.~\ref{models}).

The spectrum between these two main periodicities is close to a power-law with the slope of about $\tau^{2/3}$. This is in a accordance with findings by \citet{Plunian2009MNRAS}.
We note that a power-law scaling is typical for various convective or turbulent systems. 
It also should be verified in the framework of dynamo modelling (see  Sect.~\ref{models} below).

\begin{figure}
\centering
\includegraphics[width=\columnwidth]{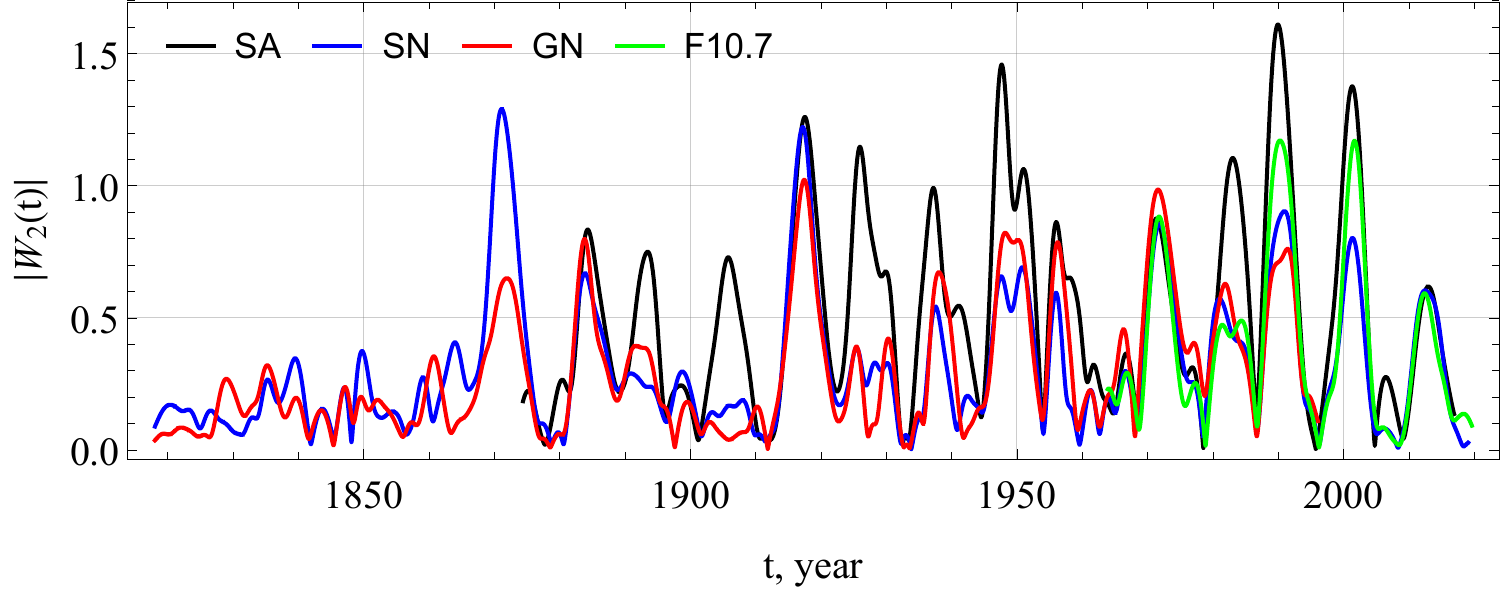}
\caption{Wavelet amplitude of the 2-year periodicity (cut off the whole wavelet spectrum) in sunspot data: SA (black), SN (blue), GN (red) and  {F10.7 (green)}.}
\label{fig:2y}
\end{figure}

The timescale of $\tau \approx 2$ years related to the QBO \citep[e.g.,][]{2014SSRv..186..359B} is not pronounced in the 
 spectra as a separate peak. 
Of course, the spectrum of sunspot data contains some power also in the QBO: Fig.~\ref{fig:2y} shows oscillations of the wavelet amplitude at the 2-year period. One can see that the 2-year oscillations substantially vary between different tracers, and also in time.  {For example, relative amplitudes differ by a factor of two during the last five solar circles, when all the considered tracers are available.} 

\begin{figure}
\centering
\includegraphics[width=\columnwidth]{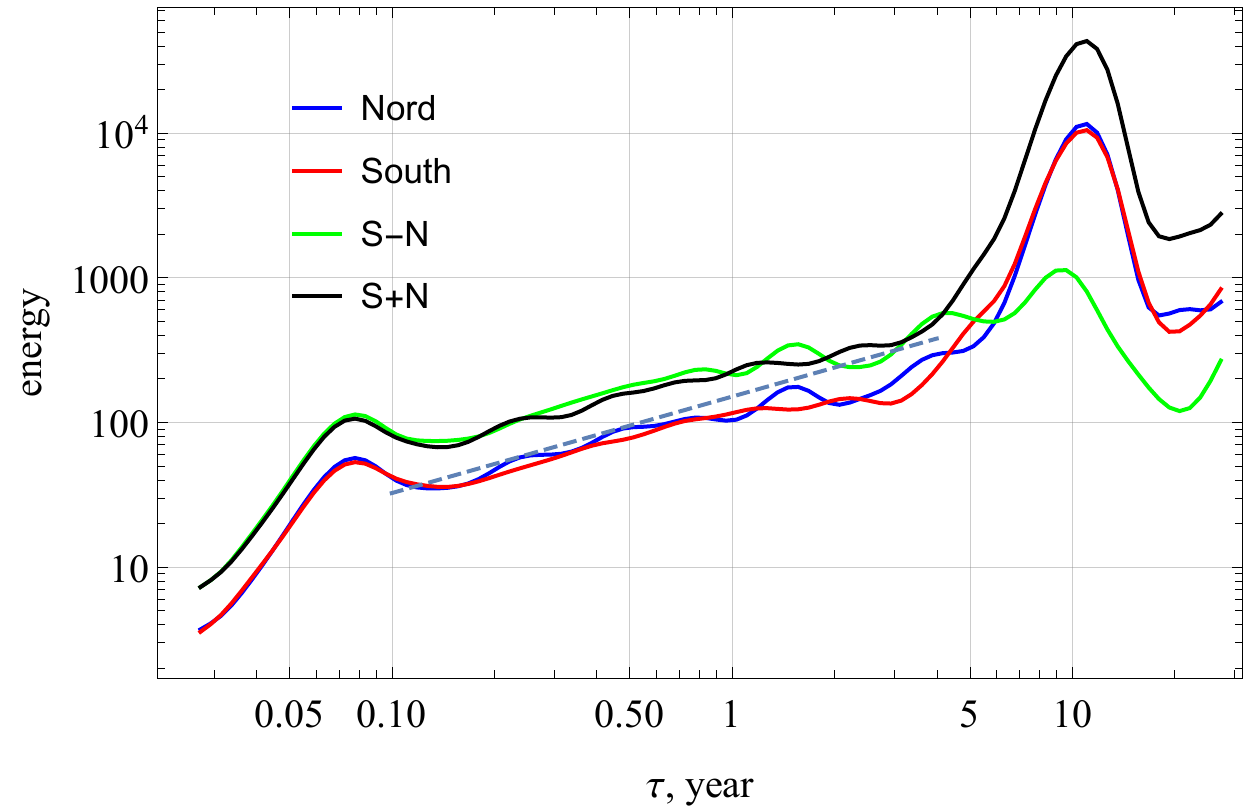}
\caption{Wavelet spectra for SA with North (blue) and South (red) hemispheres considered separately. Green line stands for spectrum of S-N data while black stands for S+N data. Dashed line corresponds to the slope ${2/3}$.}
\label{fig:specNS}
\end{figure}

Schwabe cycle is a global feature of the Sun and its convection zone. However, solar activity is not perfectly symmetric between the solar isolated hemispheres. Therefore it is useful to look for a trace of a particular periodicity in each hemisphere separately, as well as in their combination.
Thus, we separate the SA data-set in contributions from the Northern ($N$) and Southern ($S$) hemispheres and consider their sum $N+S$, i.e. the total signal, and the difference $N-S$, i.e. the excess of the sunspots area in the North over South.
Corresponding power spectral densities are shown in Fig.~\ref{fig:specNS}. The spectra of separated $N$ and $S$ data subsets are very similar, though a weak peak at $\tau\approx 1.5$ year occurs in the spectrum of the north hemisphere data. The sum $N+S$ and difference $N-S$ allow us to analyze the equatorial (anti)symmetry. Since the Schwabe cycle is produced by a global process, which is symmetric, the power of the corresponding peak in $N+S$ spectrum is expected to be quadrupole with respect to that of a single hemisphere, while in $N-S$ spectrum it would become very weak (the contributions of the North and South cancel out). 
In contrast, the power of the rotation-related peak at about 27 days is equal in both $N+S$ and $N-S$ series. It means that the corresponding variability are provided by isolated (relatively rare) long-living spots, which occur in both hemisphere independently without correlation and equatorial symmetry.
The spectral density between 1 month and 11 years behaves similar, supporting the conclusion that this range of frequency is not dominated by global phenomena presented in the Sun as a whole.

\begin{figure}
\includegraphics[width=\columnwidth,right]{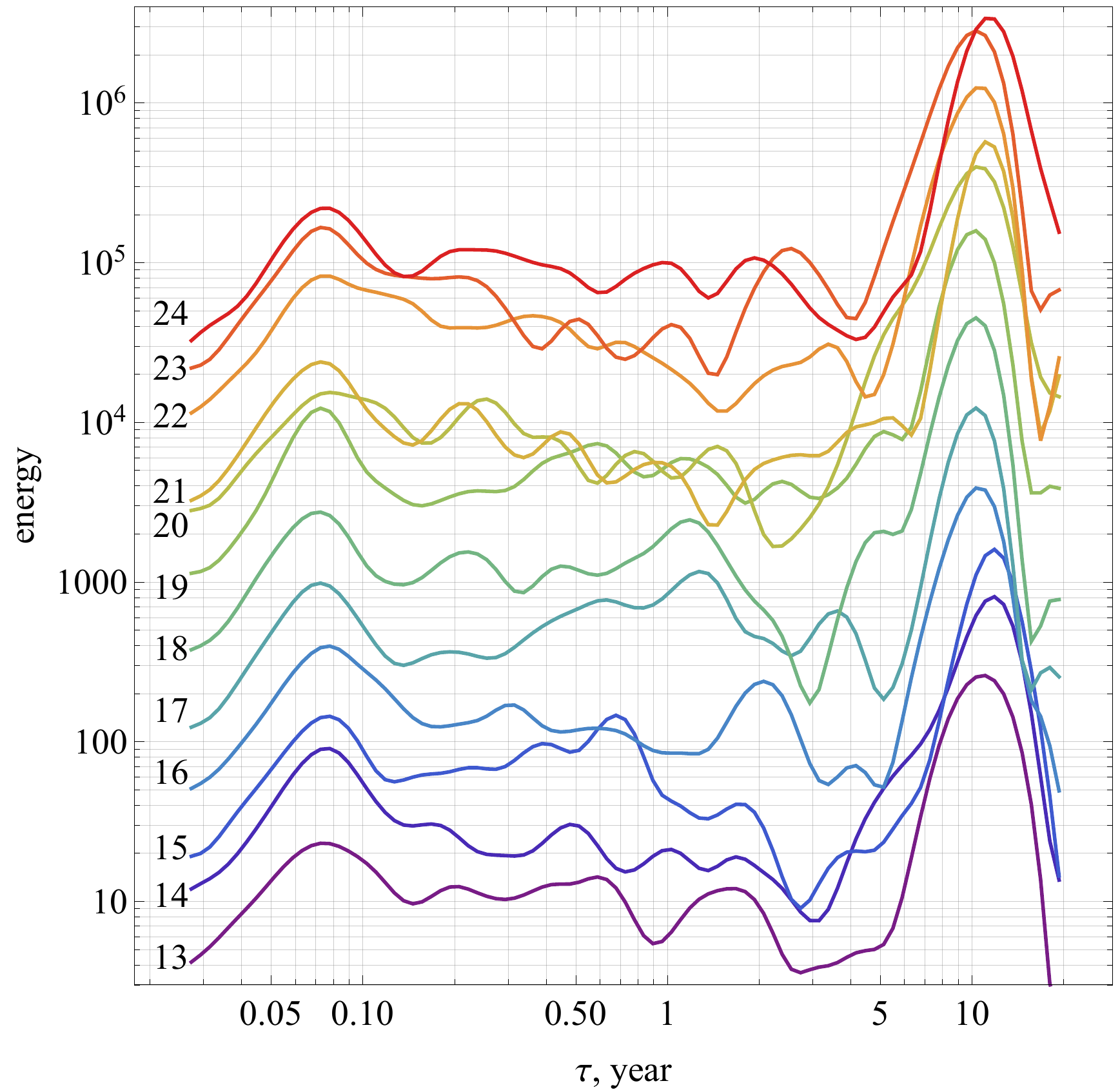}
\includegraphics[width=0.95\columnwidth,right]{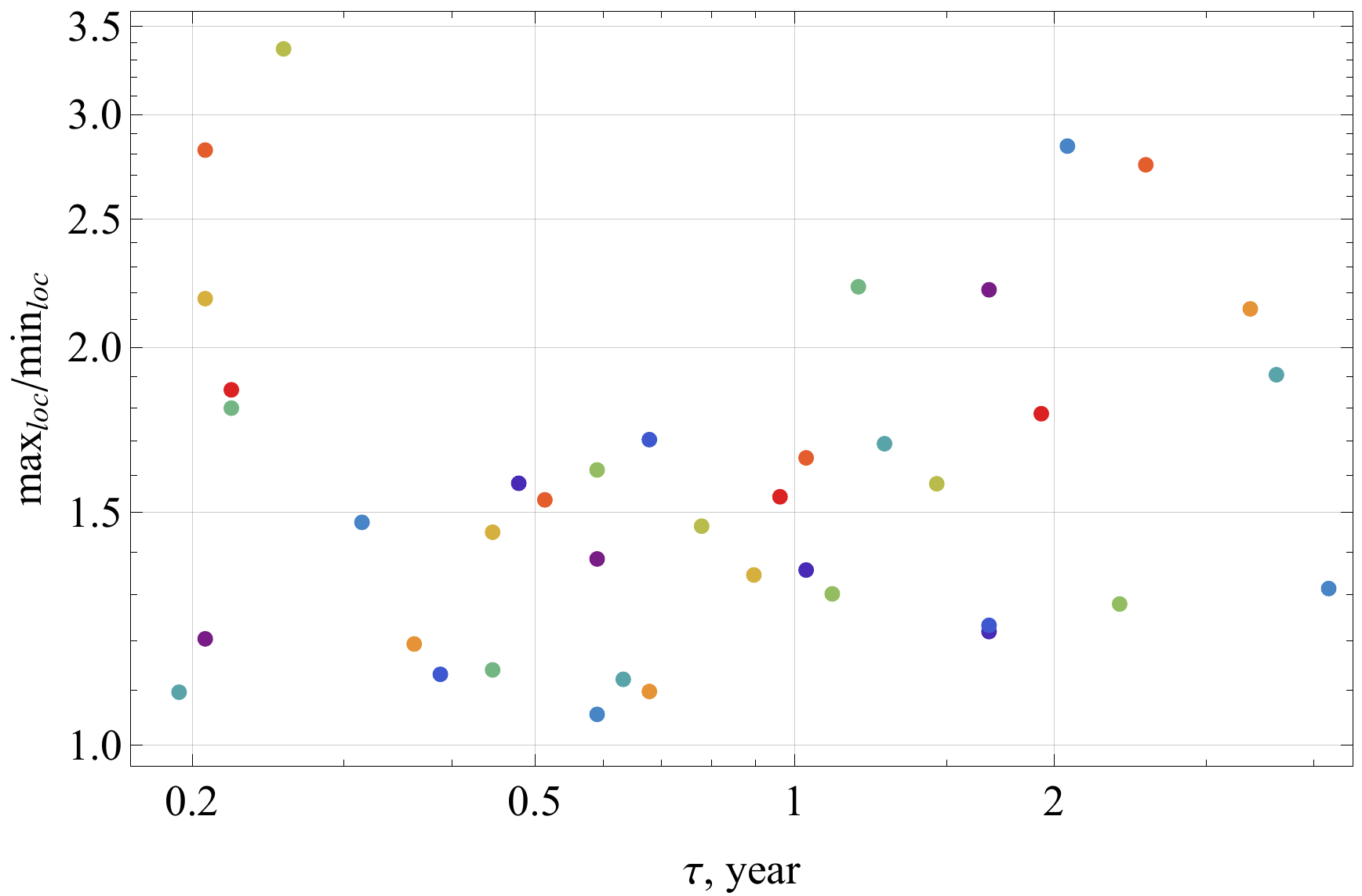}
\includegraphics[width=\columnwidth,right]{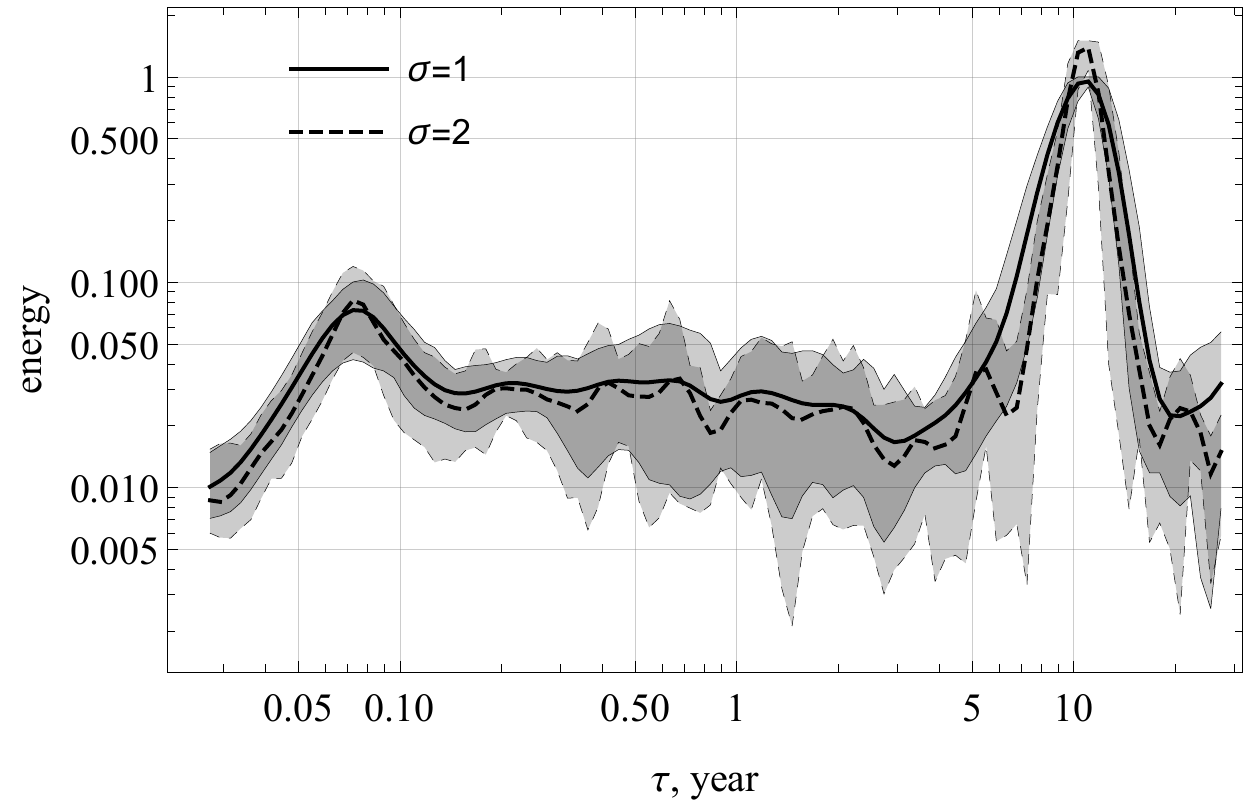}
\caption{ (top) Wavelet spectra for individual cycles, divided by $\tau^{2/3}$ and scaled by factor $10^\frac{n-13}{3}$ ($n$ is the cycle number) for better visualization. 
(middle) Local maxima relative to the nearest minima vs their periods for individual cycles.  
(bottom) The wavelet spectra averaged over individual Schwabe cycles (individual spectra were rescaled to be the same amplitude at 11 years and 
normalized by the power law $\tau^{2/3}$). 
Gray shading depicts the 80\% two-sided confidence interval. 
Solid and dashed lines correspond to different values of wavelet parameter: $\sigma=1$ and $\sigma=2$, respectively.  }
\label{fig:spec1}
\end{figure}

Supposing that some additional distinguished periodicity exists in the solar activity spectrum one should accept the requirement that corresponding oscillations must be pronounced over few solar cycles. In this case it will be captured by wavelet spectrum regardless of its appearance with random phase and amplitude.   Addressing this point we analyze the spectral power density cycle by cycle, presenting in  Fig.~\ref{fig:spec1} the wavelet spectra for 11 isolated solar cycles, calculated following Eq.~(\ref{Ea}), where $t_0$ and $t_1$ define the duration of a given cycle. One can see, that only two peaks survive over all cycles, while the spectral composition between them varies from cycle to cycle, yielding no stable features. We identified local maxima in the period range from $\tau=0.2$ to $\tau=4$. 
Middle panel of Fig.~\ref{fig:spec1}(middle) shows the relative height of local maxima found in the individual cycle spectrum, with respect to the nearest local minima. 
One can recognize two high maxima at $\tau=2$ (in cycle $N=16$) and $\tau=2.7$ (in cycle $N=23$) years. There are many others slightly less significant maxima which are uniformly occurred in mid-term range. We attempt to confirm it in a statistical manner.  Fig.~\ref{fig:spec1}(bottom) shows confidential internals around the mean values obtained by averaging over all the analyzed cycles.        
 {We can summarize the results of the wavelet analysis as that the pronounced oscillations with timescales between several months and 11 years are present in each individual cycle.} Quite naturally, however, their random contributions yields a smooth overall spectrum. 
For larger $\tau$ deviations from the simple $\tau^{2/3}$ relation become visible, and the difference from spectra obtained for individual Schwabe cycles becomes larger due to a smaller statistic. 
We understand it as possible isles of intermittent nature of solar dynamo. Strong intermittency was testified by calculating the corresponding scaling exponents in \cite{Plunian2009MNRAS}. Perhaps, the scale of $\tau \approx 2$ years can be suggested as a timescale where intermittent effects are substantially pronounced.

\section{Comparison with a solar dynamo model}
\label{models}

The observed power-law shape of the global wavelet spectra in the period range from 1 month to $\approx$11 years (Fig.~\ref{fig:specNS}) calls for an interpretation in the frameworks of a relevant dynamo processes operating inside the Sun. 
Starting from the seminal paper by \citet{Parker1955}, it is widely assumed that the surface magnetic activity of the Sun is governed by the cyclic transformation of the large-scale poloidal magnetic field into toroidal magnetic field by means of the differential rotation and the turbulent generation of the poloidal magnetic fields from  the toroidal magnetic field by the small-scale convective cyclic motions. 
An alternative option for the latter process is mirror asymmetry, which appears due to magnetic force action during sunspot emergence, the so-called Babckok-Leighton scheme, \cite{Ch14}. Meridional circulation also participates in the process. 

The turbulent part of the dynamo process is not well understood. 
Direct numerical simulations, e.g., \cite{guer16} and \citet{viv2018AA}, can reproduce the cyclic magnetic activity in a form of dynamo waves. 
However, the properties of the dynamo wave patterns in the models are different from the solar observations. 
The mean-field models can reproduce the dynamo waves of solar magnetic activity either as a result of magnetic flux transport due the meridional circulation \cite{Ch14} or as a result of diffusive dynamo waves from the dynamo distributed over convection zone \citep{ax05,pip11}. 
Those models can explain the axisymmetric components of magnetic activity but they do not take into account effect of the nonaxisymmetric magnetic fields which are produced from the solar active regions emerging and decaying on the solar surface. 

Here we apply a dynamo model of \cite{2018ApJ...867..145P}, which take this process into account using the nonaxisymmetric  $\alpha^2 \omega$ dynamo model. 
This model is formulated for the non-axisymmetric 3D magnetic field on the sphere using the shallow-water approximation \citep{dikp2001}. We use the magnetic field decomposition into toroidal and poloidal potentials and their spectral representation via the spherical harmonics. Following the shallow-water approximation it is assumed that the the poloidal part of the magnetic potential is independent of the radial coordinate and the toroidal potential is a linear function of the radius.
The dynamo is driven by the differential rotation and mirror-asymmetric convection. The effect of turbulence on evolution of the large-scale magnetic field is parameterized via the mean electromotive force \citep{KrauseRaedler:MeanFieldBook}. 
 By construction, this  model does not produce  the mid-term range oscillations. Such oscillations can be connected with the sunspot-like activity which induces the nonaxisymmetric magnetic field.  
In order to simulate the sunspot formation and induce the  nonaxisymmetric magnetic fields, the model includes the Parker buoyancy, which produces bipolar regions from the toroidal magnetic field at random latitude and time when the magnetic field strength exceeds a critical threshold. 
In this paper we roughly  assume that the sunspots production timescale is about 10 days.
More details about the model can be found in \cite{2018ApJ...867..145P}. The model yields cycles with the mean period of  $\approx 0.15\tau^*$  ($\tau^*$ being the diffusion timescale). 

Here we used data sets from the dynamo model covering a sequence  of 20 dynamo cycles. 
We have checked that longer simulation does not change the spectral property in considered range of $\tau$. 
The signal of magnetic activity produced by the bipolar regions can be used to detect the rotational period. 
To compare the results of the dynamo model with observations we calculate time series of the magnitude of the total flux of the radial  magnetic field averaged over hemisphere (longitudinal range from 0 to $\pi$).

\begin{figure}
\centering
\includegraphics[width=\columnwidth]{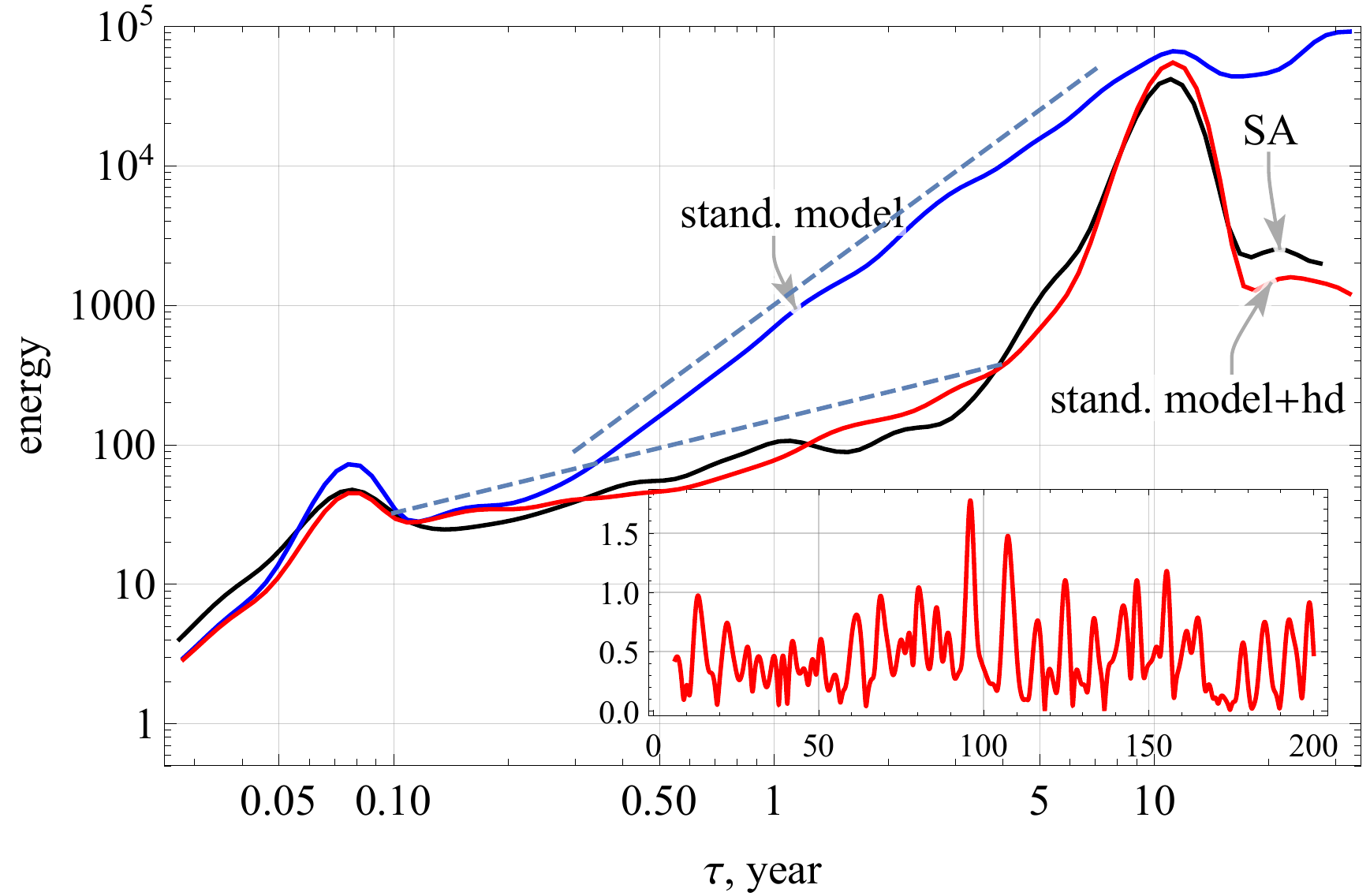}
\caption{Wavelet spectra for SN (black) and two simulation runs: (blue) standard dynamo model (see, \citealp{2018ApJ...867..145P}) and (red) this model with hyperdiffusion effect. Dashed lines stand for the slopes $2$ and $2/3$.  {Inlet figure shows wavelet amplitude evolution of the 2-year periodicity (similar to Fig.~\ref{fig:2y}) from data of mod+hd simulation.}}
\label{fig:model}
\end{figure}

Comparison between the wavelet spectra for the simulated and real data is presented in Fig.~\ref{fig:model}.
It is found that with the standard formulation of the model, i.e., the same as in \citep{2018ApJ...867..145P} the power spectrum is much steeper ($E(\tau)\sim\tau^2$ in mid-term range) and actually takes up the $11$ years peak (see the blue line labeled "model" in Fig.~\ref{fig:model}). 
We found that short-term fluctuations of the $\alpha$-effect increase the slope of the spectrum. 
Next we tested the concept of hyperdiffusion ($\Delta ^2$) for nonaxisymmetric magnetic field.  
In this case the spectrum (the red line in Fig.~\ref{fig:model}) becomes qualitatively similar to the real solar spectrum. 
We note that difference between simulated ("model+hyperdiff") and real ("SN") spectra is more remarkable near $\tau \approx 2$ years, but it stays within confidential intervals in Fig.~\ref{fig:spec1}.  { The inlet figure demonstrates that a signal similar to QBO can appear even in the output of a model, which does not contain such an intrinsic (quasi) periodicity. Irrespective of the (in)correctness of the model, this implies that the QBO (or other mid-term periodicities) can appear as a result of a random realization and/or as an artefact of the analysis method.}

It is important for our interpretation, that the  dynamo model does not contain physical mechanisms which would lead to excitation of any other particular periodicity than the Schwabe cycle. 
We manage to mimic a peak associated with the solar rotation with special treatment of simulated data. Results are in a good agreement with real data due to sufficiently long-lived surface phenomena and its homogeneous distribution over surface.
Interpretation of spectrum between these peaks looks natural in the framework of the model under discussion.  

\section{Discussion and Conclusions}

Performed detailed wavelet analysis of available sunspot data reveals only two significant periodicities in the frequency/period range between weeks  and decades, viz. the solar rotation period and the Schwabe cycle. The latter can be identified with an eigen solution of the solar dynamo, which may be saturated by some nonlinear effects.
Considering the spectral dynamics of solar activity cycle by cycle, we showed that no particular periodicity can be identified, in a statistically significant manner, in the specified range of periods. 

The spectrum in the range of periods between 1 month and 11 years presumably represents various random components of the solar dynamo system. QBO can be recognized in some individual cycles, but the available data do not make a basis to identify it with an eigen solution of solar dynamo.
This implies that there is no need to consider a specific mechanism to explain QBO and it is sufficient to consider them as  elements of continuous spectrum typical for various turbulent convective systems.  {By comparing the results obtained using different data sets for solar activity, we verified that our results are robust with respect to the choice of the data set.}

Moreover, we confirm that the spectrum in this frequency range is smooth and under reliable statistics tends to be close to the power law $E(\tau)\sim \tau^{2/3}$. This result is important because
allows us to formulate a new criterion for the verification of dynamo models pretending to describe the temporal dynamics of solar activity.  Such characteristics (the continuity of the temporal spectrum) was never been considered as a requirement for a dynamo model.  {In a recent study, \citet{2019A&A...625A..28C} have shown that oscillations with periods longer than the 11-year solar cycle appear with a overall shape of the power spectrum, which can be well represented by a generic normal form model for a noisy and weakly nonlinear limit cycle. This is also in favour of the turbulent nature of the observed solar activity apart from dominant 11-year variation.}

We examine a relatively simple dynamo model to reproduce the temporal evolution of sunspots. In the framework of the model, active regions are produced by the large-scale dynamo action. In this case nonlinear interactions between the large- and small-scale modes of magnetic field are strong. This can explain the steeper wavelet spectrum for the standard formulation of the model.
We found that a realistic spectrum of activity-related characteristic can be obtained if the small spatial (but not temporal) scales are smoothed. In the model it was realized using the concept of hyperdiffusion. 
On the Sun, the dynamo operates in the depth of the convection zone, and the surface magnetic field represents the decaying part of the dynamo wave. This result is important because the solar activity is a {\it global} feature, although recognized in small-scale tracers like sunspots. Of course, it does not imply that this model adequately reproduces all solar physics which is responsible  for sunspot statistic. However the available data do not allow to prefer the specific model over the others.

\section*{Acknowledgements}
PF, DS and RS thank the support of this work by the project RSF-Helmholtz (contracts no. 18-41-06201 and no. HRSF-0044). VP conducted study as a part of FR II.16 of ISTP SB RAS.
IU acknowledges support from the Academy of Finland (projects 307411 ReSoLVE and 321882 ESPERA). The dynamo model was tested during 
"Solar Helicities in Theory and Observations:
Implications for Space Weather and Dynamo Theory" Program at Nordic Institute for Theoretical Physics (NORDITA).

\bibliographystyle{mnras}
\bibliography{ref}

\begin{thebibliography}{}
\makeatletter
\relax
\def\mn@urlcharsother{\let\do\@makeother \do\$\do\&\do\#\do\^\do\_\do\%\do\~}
\def\mn@doi{\begingroup\mn@urlcharsother \@ifnextchar [ {\mn@doi@}
  {\mn@doi@[]}}
\def\mn@doi@[#1]#2{\def\@tempa{#1}\ifx\@tempa\@empty \href
  {http://dx.doi.org/#2} {doi:#2}\else \href {http://dx.doi.org/#2} {#1}\fi
  \endgroup}
\def\mn@eprint#1#2{\mn@eprint@#1:#2::\@nil}
\def\mn@eprint@arXiv#1{\href {http://arxiv.org/abs/#1} {{\tt arXiv:#1}}}
\def\mn@eprint@dblp#1{\href {http://dblp.uni-trier.de/rec/bibtex/#1.xml}
  {dblp:#1}}
\def\mn@eprint@#1:#2:#3:#4\@nil{\def\@tempa {#1}\def\@tempb {#2}\def\@tempc
  {#3}\ifx \@tempc \@empty \let \@tempc \@tempb \let \@tempb \@tempa \fi \ifx
  \@tempb \@empty \def\@tempb {arXiv}\fi \@ifundefined
  {mn@eprint@\@tempb}{\@tempb:\@tempc}{\expandafter \expandafter \csname
  mn@eprint@\@tempb\endcsname \expandafter{\@tempc}}}

\bibitem[\protect\citeauthoryear{{Ahlers}, {Grossmann}  \& {Lohse}}{{Ahlers}
  et~al.}{2009}]{2009RvMP...81..503A}
{Ahlers} G.,  {Grossmann} S.,   {Lohse} D.,  2009, \mn@doi [Reviews of Modern
  Physics] {10.1103/RevModPhys.81.503}, \href
  {http://adsabs.harvard.edu/abs/2009RvMP...81..503A} {81, 503}

\bibitem[\protect\citeauthoryear{{Bazilevskaya}, {Broomhall}, {Elsworth}  \&
  {Nakariakov}}{{Bazilevskaya} et~al.}{2014}]{2014SSRv..186..359B}
{Bazilevskaya} G.,  {Broomhall} A.-M.,  {Elsworth} Y.,   {Nakariakov} V.~M.,
  2014, \mn@doi [\ssr] {10.1007/s11214-014-0068-0}, \href
  {http://adsabs.harvard.edu/abs/2014SSRv..186..359B} {186, 359}

\bibitem[\protect\citeauthoryear{{Benevolenskaya}}{{Benevolenskaya}}{1995}]{1995SoPh..161....1B}
{Benevolenskaya} E.~E.,  1995, \mn@doi [\solphys] {10.1007/BF00732080}, \href
  {https://ui.adsabs.harvard.edu/abs/1995SoPh..161....1B} {161, 1}

\bibitem[\protect\citeauthoryear{{Bogatyrev}, {Gilev}  \& {Zimin}}{{Bogatyrev}
  et~al.}{1980}]{1980JETPL..32..210B}
{Bogatyrev} G.~P.,  {Gilev} V.~G.,   {Zimin} V.~D.,  1980, Soviet Journal of
  Experimental and Theoretical Physics Letters, \href
  {http://adsabs.harvard.edu/abs/1980JETPL..32..210B} {32, 210}

\bibitem[\protect\citeauthoryear{{Brandenburg}}{{Brandenburg}}{2005}]{ax05}
{Brandenburg} A.,  2005, \mn@doi [\apj] {10.1086/429584}, \href
  {https://ui.adsabs.harvard.edu/\#abs/2005ApJ...625..539B} {625, 539}

\bibitem[\protect\citeauthoryear{{Busse}}{{Busse}}{1983}]{1983PhyD....9..287B}
{Busse} F.~H.,  1983, \mn@doi [Physica D Nonlinear Phenomena]
  {10.1016/0167-2789(83)90273-7}, \href
  {http://adsabs.harvard.edu/abs/1983PhyD....9..287B} {9, 287}

\bibitem[\protect\citeauthoryear{{Cameron} \& {Sch{\"u}ssler}}{{Cameron} \&
  {Sch{\"u}ssler}}{2019}]{2019A&A...625A..28C}
{Cameron} R.~H.,  {Sch{\"u}ssler} M.,  2019, \mn@doi [\aap]
  {10.1051/0004-6361/201935290}, \href
  {https://ui.adsabs.harvard.edu/abs/2019A&A...625A..28C} {625, A28}

\bibitem[\protect\citeauthoryear{{Charbonneau}}{{Charbonneau}}{2014}]{Ch14}
{Charbonneau} P.,  2014, \mn@doi [\araa] {10.1146/annurev-astro-081913-040012},
  \href {https://ui.adsabs.harvard.edu/abs/2014ARA&A..52..251C} {52, 251}

\bibitem[\protect\citeauthoryear{{Clette} \& {Lef{\`e}vre}}{{Clette} \&
  {Lef{\`e}vre}}{2016}]{clette16}
{Clette} F.,  {Lef{\`e}vre} L.,  2016, \mn@doi [Solar Phys.]
  {10.1007/s11207-016-1014-y}, \href
  {http://adsabs.harvard.edu/abs/2016SoPh..291.2629C} {291, 2629}

\bibitem[\protect\citeauthoryear{{Dikpati} \& {Gilman}}{{Dikpati} \&
  {Gilman}}{2001}]{dikp2001}
{Dikpati} M.,  {Gilman} P.~A.,  2001, \mn@doi [\apj] {10.1086/320080}, \href
  {https://ui.adsabs.harvard.edu/abs/2001ApJ...551..536D} {551, 536}

\bibitem[\protect\citeauthoryear{{Dikpati}, {McIntosh}, {Bothun}, {Cally},
  {Ghosh}, {Gilman}  \& {Umurhan}}{{Dikpati} et~al.}{2018}]{Detal18}
{Dikpati} M.,  {McIntosh} S.~W.,  {Bothun} G.,  {Cally} P.~S.,  {Ghosh} S.~S.,
  {Gilman} P.~A.,   {Umurhan} O.~M.,  2018, \mn@doi [\apj]
  {10.3847/1538-4357/aaa70d}, \href
  {https://ui.adsabs.harvard.edu/abs/2018ApJ...853..144D} {853, 144}

\bibitem[\protect\citeauthoryear{{Fletcher}, {Broomhall}, {Salabert}, {Basu},
  {Chaplin}, {Elsworth}, {Garcia}  \& {New}}{{Fletcher} et~al.}{2010}]{Fetal10}
{Fletcher} S.~T.,  {Broomhall} A.-M.,  {Salabert} D.,  {Basu} S.,  {Chaplin}
  W.~J.,  {Elsworth} Y.,  {Garcia} R.~A.,   {New} R.,  2010, \mn@doi [\apjl]
  {10.1088/2041-8205/718/1/L19}, \href
  {https://ui.adsabs.harvard.edu/abs/2010ApJ...718L..19F} {718, L19}

\bibitem[\protect\citeauthoryear{{Frick}, {Galyagin}, {Hoyt}, {Nesme-Ribes},
  {Schatten}, {Sokoloff}  \& {Zakharov}}{{Frick} et~al.}{1997}]{Frick97}
{Frick} P.,  {Galyagin} D.,  {Hoyt} D.~V.,  {Nesme-Ribes} E.,  {Schatten}
  K.~H.,  {Sokoloff} D.,   {Zakharov} V.,  1997, \aap, \href
  {http://adsabs.harvard.edu/abs/1997A%26A...328..670F} {328, 670}

\bibitem[\protect\citeauthoryear{Frick, Grossmann  \& Tchamitchian}{Frick
  et~al.}{1998}]{Frick1998}
Frick P.,  Grossmann A.,   Tchamitchian P.,  1998, \mn@doi [J. Math. Phys.]
  {10.1063/1.532485}, 39, 4091

\bibitem[\protect\citeauthoryear{{Guerrero}, {Smolarkiewicz}, {de Gouveia Dal
  Pino}, {Kosovichev}  \& {Mansour}}{{Guerrero} et~al.}{2016}]{guer16}
{Guerrero} G.,  {Smolarkiewicz} P.~K.,  {de Gouveia Dal Pino} E.~M.,
  {Kosovichev} A.~G.,   {Mansour} N.~N.,  2016, \mn@doi [\apj]
  {10.3847/2041-8205/828/1/L3}, \href
  {https://ui.adsabs.harvard.edu/\#abs/2016ApJ...828L...3G} {828, L3}

\bibitem[\protect\citeauthoryear{Hathaway}{Hathaway}{2015}]{hathawayLR}
Hathaway D.~H.,  2015, \mn@doi [Living Rev. Solar Phys.] {10.1007/lrsp-2015-4},
  12, 4

\bibitem[\protect\citeauthoryear{{Inceoglu}, {Simoniello}, {Arlt}  \&
  {Rempel}}{{Inceoglu} et~al.}{2019}]{Ietal19}
{Inceoglu} F.,  {Simoniello} R.,  {Arlt} R.,   {Rempel} M.,  2019, \mn@doi
  [\aap] {10.1051/0004-6361/201935272}, \href
  {https://ui.adsabs.harvard.edu/abs/2019A&A...625A.117I} {625, A117}

\bibitem[\protect\citeauthoryear{Krause \& R{\"a}dler}{Krause \&
  R{\"a}dler}{1980}]{KrauseRaedler:MeanFieldBook}
Krause F.,  R{\"a}dler K.-H.,  1980, Mean-field Magnetohydrodynamics and Dynamo
  Theory.
Pergamon Press, New York

\bibitem[\protect\citeauthoryear{{Krishnamurti} \& {Howard}}{{Krishnamurti} \&
  {Howard}}{1981}]{1981PNAS...78.1981K}
{Krishnamurti} R.,  {Howard} L.~N.,  1981, \mn@doi [Proceedings of the National
  Academy of Science] {10.1073/pnas.78.4.1981}, \href
  {http://adsabs.harvard.edu/abs/1981PNAS...78.1981K} {78, 1981}

\bibitem[\protect\citeauthoryear{{Lawrence}, {Cadavid}  \&
  {Ruzmaikin}}{{Lawrence} et~al.}{1995}]{1995ApJ...455..366L}
{Lawrence} J.~K.,  {Cadavid} A.~C.,   {Ruzmaikin} A.~A.,  1995, \mn@doi [\apj]
  {10.1086/176583}, \href
  {https://ui.adsabs.harvard.edu/abs/1995ApJ...455..366L} {455, 366}

\bibitem[\protect\citeauthoryear{Mallat}{Mallat}{2008}]{Mallat2008}
Mallat S.,  2008, A Wavelet Tour of Signal Processing, Third Edition: The
  Sparse Way, 3rd edn.
Academic Press

\bibitem[\protect\citeauthoryear{{Nesme-Ribes}, {Frick}, {Sokoloff},
  {Zakharov}, {Ribes}, {Vigouroux}  \& {Laclare}}{{Nesme-Ribes}
  et~al.}{1995}]{1995CRASB.321..525N}
{Nesme-Ribes} E.,  {Frick} P.,  {Sokoloff} D.,  {Zakharov} V.,  {Ribes} J.~C.,
  {Vigouroux} A.,   {Laclare} F.,  1995, Academie des Sciences Paris Comptes
  Rendus Serie B Sciences Physiques, \href
  {https://ui.adsabs.harvard.edu/abs/1995CRASB.321..525N} {12, 525}

\bibitem[\protect\citeauthoryear{{Niemela}, {Skrbek}, {Sreenivasan}  \&
  {Donnelly}}{{Niemela} et~al.}{2001}]{2001JFM...449..169N}
{Niemela} J.~J.,  {Skrbek} L.,  {Sreenivasan} K.~R.,   {Donnelly} R.~J.,  2001,
  \mn@doi [Journal of Fluid Mechanics] {10.1017/S0022112001006310}, \href
  {http://adsabs.harvard.edu/abs/2001JFM...449..169N} {449, 169}

\bibitem[\protect\citeauthoryear{{Parker}}{{Parker}}{1955}]{Parker1955}
{Parker} E.~N.,  1955, \mn@doi [\apj] {10.1086/146087}, \href
  {http://adsabs.harvard.edu/abs/1955ApJ...122..293P} {122, 293}

\bibitem[\protect\citeauthoryear{{Pipin} \& {Kosovichev}}{{Pipin} \&
  {Kosovichev}}{2011}]{pip11}
{Pipin} V.~V.,  {Kosovichev} A.~G.,  2011, \mn@doi [\apj]
  {10.1088/0004-637X/741/1/1}, \href
  {https://ui.adsabs.harvard.edu/\#abs/2011ApJ...741....1P} {741, 1}

\bibitem[\protect\citeauthoryear{{Pipin} \& {Kosovichev}}{{Pipin} \&
  {Kosovichev}}{2018}]{2018ApJ...867..145P}
{Pipin} V.~V.,  {Kosovichev} A.~G.,  2018, \mn@doi [\apj]
  {10.3847/1538-4357/aae1fb}, \href
  {https://ui.adsabs.harvard.edu/\#abs/2018ApJ...867..145P} {867, 145}

\bibitem[\protect\citeauthoryear{{Plunian}, {Sarson}  \& {Stepanov}}{{Plunian}
  et~al.}{2009}]{Plunian2009MNRAS}
{Plunian} F.,  {Sarson} G.~R.,   {Stepanov} R.,  2009, \mn@doi [\mnras]
  {10.1111/j.1745-3933.2009.00760.x}, \href
  {http://adsabs.harvard.edu/abs/2009MNRAS.400L..47P} {400, L47}

\bibitem[\protect\citeauthoryear{{Qiu} \& {Tong}}{{Qiu} \&
  {Tong}}{2001}]{2001PhRvL..87i4501Q}
{Qiu} X.-L.,  {Tong} P.,  2001, \mn@doi [Physical Review Letters]
  {10.1103/PhysRevLett.87.094501}, \href
  {http://adsabs.harvard.edu/abs/2001PhRvL..87i4501Q} {87, 094501}

\bibitem[\protect\citeauthoryear{{Simoniello}, {Jain}, {Tripathy},
  {Turck-Chi{\`e}ze}, {Baldner}, {Finsterle}, {Hill}  \& {Roth}}{{Simoniello}
  et~al.}{2013}]{Setal13}
{Simoniello} R.,  {Jain} K.,  {Tripathy} S.~C.,  {Turck-Chi{\`e}ze} S.,
  {Baldner} C.,  {Finsterle} W.,  {Hill} F.,   {Roth} M.,  2013, \mn@doi [\apj]
  {10.1088/0004-637X/765/2/100}, \href
  {https://ui.adsabs.harvard.edu/abs/2013ApJ...765..100S} {765, 100}

\bibitem[\protect\citeauthoryear{{Soon}, {Frick}  \& {Baliunas}}{{Soon}
  et~al.}{1999}]{1999ApJ...510L.135S}
{Soon} W.,  {Frick} P.,   {Baliunas} S.,  1999, \mn@doi [\apjl]
  {10.1086/311805}, \href
  {https://ui.adsabs.harvard.edu/abs/1999ApJ...510L.135S} {510, L135}

\bibitem[\protect\citeauthoryear{{Usoskin}}{{Usoskin}}{2017}]{usoskin_LR_17}
{Usoskin} I.~G.,  2017, \mn@doi [Living Rev. Solar Phys.]
  {10.1007/s41116-017-0006-9}, 14, 3

\bibitem[\protect\citeauthoryear{{Usoskin}, {Kovaltsov}, {Lockwood}, {Mursula},
  {Owens}  \& {Solanki}}{{Usoskin} et~al.}{2016}]{usoskin16}
{Usoskin} I.~G.,  {Kovaltsov} G.~A.,  {Lockwood} M.,  {Mursula} K.,  {Owens}
  M.,   {Solanki} S.~K.,  2016, \mn@doi [Solar Phys.]
  {10.1007/s11207-015-0838-1}, \href
  {http://adsabs.harvard.edu/abs/2016SoPh..tmp....6U} {}

\bibitem[\protect\citeauthoryear{Vaquero et~al.,}{Vaquero
  et~al.}{2016}]{vaquero16}
Vaquero J.,  et~al., 2016, \mn@doi [Solar Phys.] {10.1007/s11207-016-0982-2},
  291, 3061

\bibitem[\protect\citeauthoryear{Vasiliev, Sukhanovskii, Frick, Budnikov,
  Fomichev, Bolshukhin  \& Romanov}{Vasiliev et~al.}{2016}]{Vasiliev:IJHMT2016}
Vasiliev A.,  Sukhanovskii A.,  Frick P.,  Budnikov A.,  Fomichev V.,
  Bolshukhin M.,   Romanov R.,  2016, Int. J. Heat Mass Transfer, 102, 201

\bibitem[\protect\citeauthoryear{{Viviani}, {Warnecke}, {K{\"a}pyl{\"a}},
  {K{\"a}pyl{\"a}}, {Olspert}, {Cole-Kodikara}, {Lehtinen}  \&
  {Brandenburg}}{{Viviani} et~al.}{2018}]{viv2018AA}
{Viviani} M.,  {Warnecke} J.,  {K{\"a}pyl{\"a}} M.~J.,  {K{\"a}pyl{\"a}} P.~J.,
   {Olspert} N.,  {Cole-Kodikara} E.~M.,  {Lehtinen} J.~J.,   {Brandenburg} A.,
   2018, \mn@doi [\aap] {10.1051/0004-6361/201732191}, \href
  {https://ui.adsabs.harvard.edu/\#abs/2018A&A...616A.160V} {616, A160}

\bibitem[\protect\citeauthoryear{{Zaqarashvili}, {Carbonell}, {Oliver}  \&
  {Ballester}}{{Zaqarashvili} et~al.}{2010}]{Zetal10}
{Zaqarashvili} T.~V.,  {Carbonell} M.,  {Oliver} R.,   {Ballester} J.~L.,
  2010, \mn@doi [\apjl] {10.1088/2041-8205/724/1/L95}, \href
  {https://ui.adsabs.harvard.edu/abs/2010ApJ...724L..95Z} {724, L95}

\bibitem[\protect\citeauthoryear{{Zimin} \& {Frick}}{{Zimin} \&
  {Frick}}{1988}]{Zimin1988b}
{Zimin} V.~D.,  {Frick} P.~G.,  1988, Turbulent Convection.
Nauka, Moscow

\makeatother
\end{thebibliography}
\label{lastpage}
\end{document}